\documentclass{jfm}
\usepackage{changes}
\usepackage{graphicx}
\usepackage{newtxtext}
\usepackage{newtxmath}
\usepackage{natbib}
\usepackage{hyperref}
\usepackage{color}
\hypersetup{
    colorlinks = true,
    urlcolor   = blue,
    citecolor  = black,
}

\newcommand{\RomanNumeralCaps}[1]
\def \d{\mathrm{d}}

\title{Scaling laws for planetary sediment transport from DEM-RANS numerical simulations}

\author{Thomas P\"ahtz\aff{1,2}
	\corresp{\email{0012136@zju.edu.cn}}
  \and Orencio Dur\'an\aff{3}}

\affiliation{\aff{1}Donghai Laboratory, 316021 Zhoushan, PR China
\aff{2}Institute of Port, Coastal and Offshore Engineering, Ocean College, Zhejiang University, 316021 Zhoushan, PR China
\aff{3}Department of Ocean Engineering, Texas A\&M University, College Station, Texas 77843-3136, USA}

\begin{document}
\maketitle

\begin{abstract}
We use an established discrete element method (DEM) Reynolds-averaged Navier--Stokes (RANS)-based numerical model to simulate non-suspended sediment transport across conditions encompassing almost seven orders of magnitude in the particle--fluid density ratio $s$, ranging from subaqueous transport ($s=2.65$) to aeolian transport in the highly rarefied atmosphere of Pluto ($s=10^7$), whereas previous DEM-based sediment transport studies did not exceed terrestrial aeolian conditions ($s\approx2000$). Guided by these simulations and by experiments, we semi-empirically derive simple scaling laws for the cessation threshold and rate of equilibrium aeolian transport, both exhibiting a rather unusual $s^{1/3}$-dependence. They constitute a simple means to make predictions of aeolian processes across a large range of planetary conditions. The derivation consists of a first-principle-based proof of the statement that, under relatively mild assumptions, the cessation threshold physics is controlled by only one dimensionless control parameter, rather than two expected from dimensional analysis. Crucially, unlike existing models, this proof does not resort to coarse-graining the particle phase of the aeolian transport layer above the bed surface. From the pool of existing models, only that by P\"ahtz et al. (\textit{J. Geophys. Res. Earth. Surf.}~126, e2020JF005859, 2021) is somewhat consistent with the combined numerical and experimental data. It captures the scaling of the cessation threshold and the $s^{1/3}$-dependence of the transport rate, but fails to capture the latter's superimposed grain size dependence. This hints at a lack of understanding of the transport rate physics and calls for future studies on this issue.
\end{abstract}


\section{Introduction}
Aeolian (wind-driven) transport of non-suspended grains, including sand, ice and snow, is a ubiquitous phenomenon that leads to a rich variety of multiscale bedforms on Earth and other planetary bodies \citep{Bourkeetal10,Koketal12,Diniegaetal17}. As suggested by the presence of wind streaks and dunes, it may even occur in the very rarefied atmospheres of Neptune's moon Triton \citep{SaganChyba90}, Pluto \citep{Telferetal18} and the comet 67P/Churyumov-Gerasimenko \citep{Thomasetal15a,Jiaetal17}.

Driven by fluid drag and gravity, most transported sand-sized and larger grains regularly interact with the bed surface as flow turbulence is too weak to suspend them. For denser fluids, such as water and most other liquids, this near-surface grain motion occurs in the form of rolling, sliding and small hops (\textit{bedload}), whereas for lighter fluids, like most gases, grains move in more energetic hops (\textit{saltation}). At equilibrium, the deposition of transported grains on the bed is exactly balanced by the entrainment of bed grains into the transport layer. The rate at which equilibrium aeolian transport takes place and the threshold wind speed below which it ceases constitute the two arguably most important statistical transport properties in the context of bedform formation and evolution in natural environments \citep{Kok10a,Duranetal19}. In particular, in natural environments, topography inhomogeneities, strong turbulent fluctuations and a variety of wind-unrelated mechanisms to generate airborne grains, along with very long natural sediment fetches, can plausibly initiate transport and lead to equilibrium transport above the cessation threshold \citep[][section~3.3.3.4]{Pahtzetal20a}. This may even be true in environments where the aeolian transport initiation threshold for an idealised flat sediment bed is much larger than the cessation threshold, like potentially on Mars \citep{Kok10a}, Pluto \citep{Telferetal18} and Saturn's moon Titan \citep{Comolaetal22}, as well as in Antarctica. In fact, although Antarctica's surface is covered by very cohesive \citep{PomeroyGray90} old snow and ice (cohesion increases the initiation threshold probably much more than the cessation threshold \citep{Comolaetal19b,Comolaetal22,Pahtzetal21,Besnardetal22}), aeolian snow and ice transport occurs there even at relatively low wind speeds that are likely much below the initiation threshold \citep{Leonardetal11}.

Since the highly random, collective motion of bed and transported grains eludes a rigorous analytical description, existing physical models of equilibrium aeolian transport have relied on drastically coarse-graining the particle phase of the aeolian transport layer above the bed surface \citep{UngarHaff87,Andreotti04,ClaudinAndreotti06,KokRenno09,Kok10b,Duranetal11,Berzietal16,Berzietal17,LammelKroy17,PahtzDuran18a,PahtzDuran20,Andreottietal21,Pahtzetal21,Comolaetal22,GunnJerolmack22}. The most common modelling approach is to represent the grain motion by a single or multiple saltation trajectories. Depending on the number and kind of considered trajectories and the assumed outcome of grain--bed collisions, such models can yield fundamentally different scaling laws for the cessation threshold and/or equilibrium transport rate, with predictions varying by about an order of magnitude when applied to Martian-pressure atmospheric conditions \citep{Pahtzetal20a,GunnJerolmack22}.

One reason for the strong variability of both existing cessation threshold and equilibrium transport rate predictions is a lack of consensus on the physical picture behind the cessation threshold. On the one hand, it has been modelled as an `impact entrainment threshold' \citep{Pahtzetal20a}, the smallest wind velocity at which random captures of saltating grains by the bed can be compensated by the splash of bed grains due to grain--bed impacts \citep{Andreotti04,ClaudinAndreotti06,KokRenno09,Kok10b,Andreottietal21,Comolaetal22}. On the other hand, it has been modelled as a `rebound threshold' \citep{Pahtzetal20a}, the smallest wind velocity required to replenish the energy saltating grains lose when rebounding with the bed, independent of grain capture and splash \citep{Berzietal17,Pahtzetal21,GunnJerolmack22}. We previously proposed and supported the hypothesis that both these dynamic thresholds play a role in saltation dynamics: the former as the dynamic threshold of continuous and the latter as the dynamic threshold of intermittent saltation and therefore as the actual cessation threshold \citep{PahtzDuran18a,Pahtzetal20a,Pahtzetal21}. If true, this could have the unintended consequence that measurements of one are mistaken for the other dynamic threshold. For example, \citet{Pahtzetal21} proposed that the recent dynamic-threshold measurements in a low-pressure wind tunnel by \citet{Andreottietal21} may constitute data of the continuous-transport threshold, and not of the cessation threshold as the experimenters claimed. This would be problematic as these data have been used to develop new cessation threshold models and compare their predictive capabilities with those of older ones \citep{Andreottietal21,GunnJerolmack22}.

Here, we show that, under relatively mild assumptions, one can obtain insights into the physics of the cessation threshold without resorting to coarse-graining the particle phase of the aeolian transport layer above the bed surface. In detail, if the bed surface can be considered as a flat boundary, with scale-free boundary conditions describing the outcome of grain--bed collisions, and the driving wind as a smooth inner turbulent boundary layer flow that interacts with grains via Stokes drag, then the threshold shear velocity, appropriately non-dimensionalised, is a function of only one dimensionless control parameter, rather than two expected from dimensional analysis (section~\ref{Results}). We confirm this prediction, and therefore its underlying assumptions, with numerical simulations using an existing discrete element method (DEM)-based numerical model \citep[][introduced in section~\ref{NumericalModel}]{Duranetal12} of equilibrium transport of cohesionless non-suspended sediments. The simulated transport conditions encompass almost seven orders of magnitude in the particle--fluid density ratio $s$, ranging from subaqueous transport ($s=2.65$) to aeolian transport in the highly rarefied atmosphere of Pluto ($s=10^7$), whereas previous DEM-based sediment transport studies did not exceed terrestrial aeolian conditions ($s\approx2000$). We also use the simulation data to semi-empirically derive simple scaling laws for the cessation threshold and equilibrium transport rate, and to test existing models (section~\ref{Results}). The derived scaling laws are consistent with experimental data, except the dynamic-threshold measurements by \citet{Andreottietal21}, in line with the aforementioned hypothesis that the latter constitute data of the continuous-transport threshold rather than the cessation threshold (discussed in more detail in section~\ref{DynamicThresholdMeasurements}).

\section{Numerical model} \label{NumericalModel}
We use the numerical model of \citet{Duranetal12}, which couples a continuum Reynolds-averaged description of hydrodynamics with a DEM for the grain motion under gravity, buoyancy and fluid drag. The drag force is given by $\boldsymbol{F_d}=\frac{1}{8}\rho_f\pi d^2C_d|\boldsymbol{u_r}|\boldsymbol{u_r}$, where $\rho_f$ is the fluid density, $d$ the median grain diameter, $\boldsymbol{u_r}$ the fluid--grain velocity difference and
\begin{equation}
 C_d=\left(\sqrt{\frac{\Rey_c}{|\boldsymbol{u_r}|d/\nu}}+\sqrt{C_d^\infty}\right)^2 \label{Drag}
\end{equation}
the drag coefficient, with $\nu$ the kinematic viscosity. Most simulations are carried out using the parameter values $\Rey_c=24$ and $C_d^\infty=0.5$, close to those for spherical grains \citep{Camenen07}, while a few simulations are carried out using different values (specified when done so) to test the effect of drag modifications, which may for example occur in very-low-pressure atmospheres due to drag rarefaction \citep{Croweetal12}. Spherical grains ($10^4{-}10^5$) with mild polydispersity are confined in a quasi-two-dimensional domain of length $\approx10^3d$, with periodic boundary conditions in the flow direction, and interact via normal repulsion (restitution coefficient $e=0.9$) and tangential friction (contact friction coefficient $\mu_c=0.5$). The bottom-most grain layer is glued on a bottom wall, while the top of the simulation domain is reflective but so high that it is never reached by transported grains. The Reynolds-averaged Navier--Stokes (RANS) equations are combined with a semi-empirical mixing length closure that accounts for the viscous sublayer of the turbulent boundary layer and ensures a smooth hydrodynamic transition from high to low particle concentration at the bed surface:
\begin{equation}
 \frac{\d l_m}{\d z}=\kappa\left[1-\exp\left(-\sqrt{\frac{u_xl_m}{7\nu}}\right)\right], \label{lm}
\end{equation}
where $l_m(z)$ is the height-dependent mixing length, $\kappa=0.4$ the von K\'arm\'an constant and $u_x(z)$ the mean flow velocity field. This parametrisation quantitatively reproduces measurements of $u_x(z)$ in the absence of transport. Simulations with this numerical model are insensitive to $e$ and, therefore, insensitive to viscous damping \citep{PahtzDuran18a,PahtzDuran18b}. The simulations reproduce measurements of the rate and cessation threshold of terrestrial aeolian transport, and viscous and turbulent subaqueous transport (figures~1 and 3 of \citet{PahtzDuran18a} and figure~4 of \citet{PahtzDuran20}), height profiles of relevant equilibrium transport properties (figure~2 of \citet{PahtzDuran18a} and figure~6 of \citet{Duranetal14a}) and aeolian ripple formation \citep{Duranetal14b}.

\subsection{Average of simulated quantities} \label{Averages}
We define two types of averages of a particle property $A_p$. Based on the spatial homogeneity of the simulations, the mass-weighted average of $A_p$ over all particles within an infinitesimal vertical layer $(z,z+\d z)$ and all time steps (after reaching the steady state) is \citep{PahtzDuran18b}
\begin{equation}
	\langle A\rangle(z)=\sum_{z_p\in(z,z+\d z)}m_pA_p/\sum_{z_p\in(z,z+\d z)}m_p,
\end{equation}
where $m_p$ and $z_p$ are the particle mass and elevation, respectively. We also define the average of a vertical profile $\langle A\rangle(z)$ over the transport layer as \citep{PahtzDuran18a}
\begin{equation}
 \overline{A}=\int_0^\infty\rho\langle A\rangle\d z/\int_0^\infty\rho\d z, 
 \label{TransportAverage}
\end{equation}
where $\rho$ is the local particle concentration. The bed surface elevation $z=0$ is defined as the elevation at which $p_g\d\langle v_x\rangle/\d z$ is maximal \citep{PahtzDuran18b}, where $\langle v_x\rangle$ is the average grain velocity in the streamwise direction and $p_g(z)=-\int_z^\infty\rho\langle a_z\rangle\d z^\prime$ the normal-bed granular pressure, with $\boldsymbol{a}$ the acceleration of grains by non-contact forces.

\subsection{Calculation of transport rate and cessation threshold} \label{ImportantQuantities}
We calculate the sediment transport rate $Q$ as \citep{PahtzDuran18b}
\begin{equation}
 Q=\int_{-\infty}^\infty\rho\langle v_x\rangle\d z.
\end{equation}
When $Q$ vanishes, the grain-borne shear stress at the bed surface $\tau_g(0)$ also vanishes, with $\tau_g(z)=\int_z^\infty\rho\langle a_x\rangle\d z^\prime$ the grain-borne shear stress profile. We therefore extrapolate the cessation threshold value $\tau_t$ of the fluid shear stress $\tau$ at which $Q$ vanishes using the approximate relation \citep{PahtzDuran18b}
\begin{equation}
 \tau_g(0)=\tau-\tau_t, \label{taug}
\end{equation}
where we treat $\tau_t$ as a fit parameter.

\subsection{Dimensionless control parameters and rescaling of physical quantities}
The average properties of equilibrium sediment transport are mainly determined by a few grain and environmental parameters: the grain and fluid density ($\rho_p$ and $\rho_f$, respectively), median grain diameter ($d$), kinematic fluid viscosity ($\nu$), fluid shear velocity ($u_\ast\equiv\sqrt{\tau/\rho_f}$) and gravitational constant ($g$) or its buoyancy-reduced value $\tilde g\equiv(1-\rho_f/\rho_p)g$ (for air, $\tilde g\simeq g$). Physical quantities with a superscript `+' are rescaled using units of $\rho_p$, $\tilde g$ and $\nu$. For example,
\begin{align}
 d^+&=\tilde gd/(\tilde g\nu)^{2/3}, \\
 u^+_\ast&=u_\ast/(\tilde g\nu)^{1/3}, \\
 Q^+&=Q/(\rho_p\nu).
\end{align}
As we show, this rescaling is well suited to describe the relevant physical processes underlying the cessation threshold scaling. A given environmental condition is fully determined by the values of three dimensionless numbers \citep{PahtzDuran20}:
\begin{alignat}{2}
 s&\equiv\rho_p/\rho_f&&=1/\rho_f^{+}, \\
 Ga&\equiv\sqrt{s\tilde gd^3}/\nu&&=\sqrt{s}d^{+3/2}, \\
 \Theta&\equiv u_\ast^2/(s\tilde gd)&&=u_\ast^{+2}/(sd^+).
\end{alignat}
Numerical simulations are carried out for various combinations of the particle--fluid density ratio $s$ and Galileo number $Ga$, exceeding previously simulated conditions by almost four orders of magnitude in $s$ (table~\ref{SimulationPhaseSpace}), and for Shields numbers $\Theta$ ranging from weak conditions near its cessation threshold value $\Theta_t$ to intense conditions far above $\Theta_t$.
\begin{table}
  \begin{center}
\def~{\hphantom{0}}
  \begin{tabular}{r|l}
       $s\quad$ & $\quad Ga$ \\[3pt]
       $2.65\quad$ & $\quad[0.1^\dagger,0.5^\dagger,2^\dagger,5^\dagger,10^\dagger,20^\dagger,50^\dagger,100^\dagger]$ \\
       $1\times10^1\quad$ & $\quad[20,50]$ \\
       $2\times10^1\quad$ & $\quad[20,50,100]$ \\
       $5\times10^1\quad$ & $\quad[2,5,10,20,50,100]$ \\
       $1\times10^2\quad$ & $\quad[0.1^\dagger,0.5^\dagger,2^\dagger,5^\dagger,10^\dagger,20^\dagger,50^\dagger,100^\dagger]$ \\
			 $2\times10^2\quad$ & $\quad20$ \\
			 $5\times10^2\quad$ & $\quad[2,5,10,20,50,100]$ \\
			 $1\times10^3\quad$ & $\quad10$ \\
			 $2\times10^3\quad$ & $\quad[0.1^\dagger,0.5^\dagger,1,1.5,1.7,1.8,2^\dagger,5^\dagger,10^\dagger,20^\dagger,50^\dagger,100^\dagger]$ \\
			 $5\times10^3\quad$ & $\quad[2,5]$ \\
			 $1\times10^4\quad$ & $\quad[2,5]$ \\
			 $2\times10^4\quad$ & $\quad[1,2,5]$ \\
			 $5\times10^4\quad$ & $\quad[0.1,0.5,2,3,5,10,20,50,100]$ \\
			 $2.5\times10^5\quad$ & $\quad1$ (simulations with larger $Ga$ are unstable$^\ast$) \\
			 $1\times10^7\quad$ & $\quad0.2$ (simulations with larger $Ga$ are unstable$^\ast$) \\			 
  \end{tabular}
  \caption{Simulated particle--fluid-density ratios $s$ and Galileo numbers $Ga$. $^\ast$The condition $s=2.5\times10^5$, $Ga=1$ corresponds to a typical transport environment on Mars ($d\approx100~\mathrm{\mu m}$) and $s=10^7$, $Ga=0.2$ to a hypothetical transport environment on Pluto ($d\approx200~\mathrm{\mu m}$). Simulations with significantly larger respective values of $Ga$ are unstable for these large-$s$ conditions. We have been unable to fix this issue and do not know whether it has numerical or physical causes. The asterisk symbol, $\dagger$, indicates conditions simulated in our previous studies \citep{PahtzDuran18a,PahtzDuran20}.}
  \label{SimulationPhaseSpace}
  \end{center}
\end{table}

\subsection{Sediment transport regimes for near-threshold conditions}
\begin{figure}
\centering
 \centerline{\includegraphics[width=\textwidth]{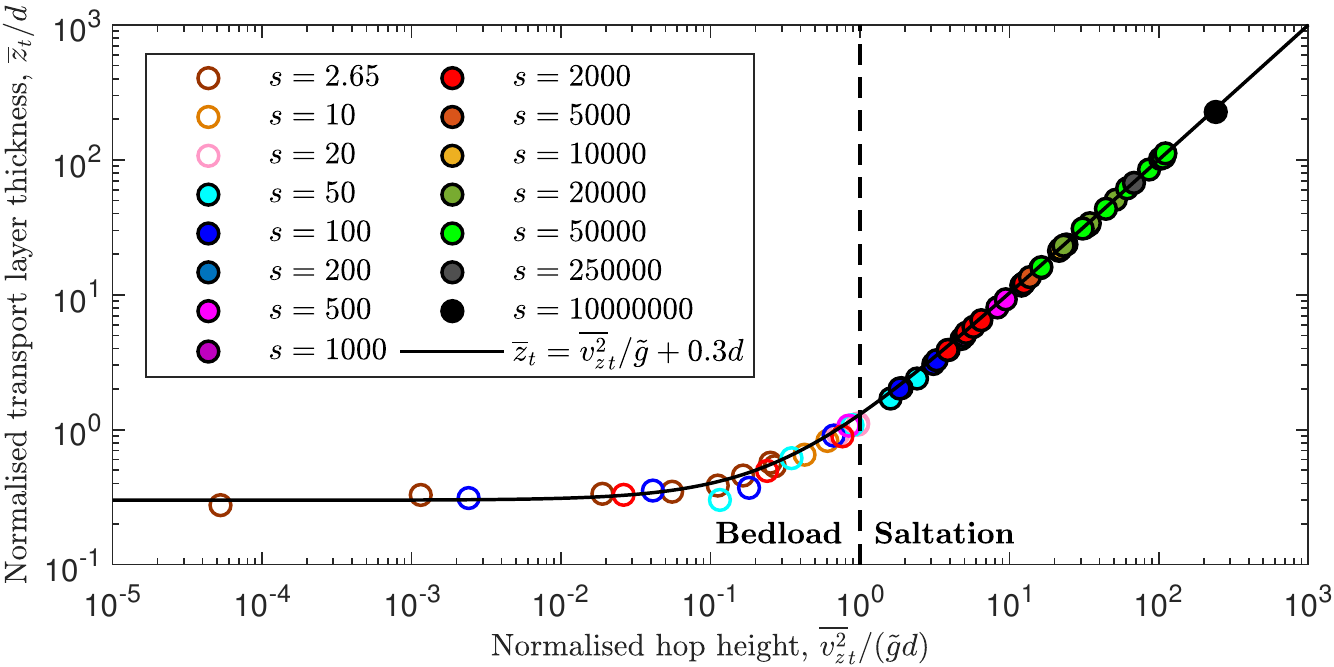}}
\caption{Transport layer thickness $\overline{z}_t$ versus hop height $\overline{v_z^2}_t/\tilde g$, both relative to the grain size $d$. Symbols correspond to numerical simulations near the cessation threshold for various combinations of the density ratio $s$ and Galileo number $Ga$ (see table~\ref{SimulationPhaseSpace}), with open and filled symbols indicating bedload and saltation conditions, respectively.}
\label{PhaseSpace}
\end{figure}

Since the mixing length-based Reynolds-averaged description of hydrodynamics used in the numerical model neglects turbulent fluctuations around the mean turbulent flow, simulated sediment transport is always non-suspended. Near the cessation threshold (subscript $t$), non-suspended transport occurs as either bedload or saltation (see the introduction), which we distinguish through the criterion \citep{PahtzDuran18a}
\begin{equation}
 \text{Transport regime}=
 \begin{cases}
  \text{bedload}&\text{if}\quad\overline{v_z^2}_t/\tilde g<d \\
	\text{saltation}&\text{if}\quad\overline{v_z^2}_t/\tilde g\geq d
 \end{cases}.
\end{equation}
The quantity $\overline{v_z^2}/\tilde g$ describes the contribution of hopping grains to the characteristic transport height of all transported grains $\overline{z}$, where the latter also include those that role and slide. In particular, for saltation near the cessation threshold, $\overline{v_z^2}_t/\tilde g\simeq\overline{z}_t$, whereas $\overline{v_z^2}_t/\tilde g$ is significantly smaller than $\overline{z}_t$ for bedload transport (figure~\ref{PhaseSpace}).
Henceforth, $\overline{v_z^2}/\tilde g$ and $\overline{z}$ are termed \textit{hop height} and \textit{transport layer thickness}, respectively, for simplicity.

\section{Results} \label{Results}
This section is organised as follows. First, it shows the data and scaling laws of the cessation threshold and equilibrium transport rate obtained from the simulations for the saltation regime (section~\ref{SaltationScalingLaws}). Second, it presents semi-empirical physical justifications of these laws, including a first-principle-based proof of the statement that, under relatively mild assumptions, the rescaled cessation threshold $u_{\ast t}^+$ is a function of only one dimensionless control parameter (section~\ref{PhysicalJustification}). Third, it tests existing models from the literature against the numerical data (section~\ref{ModelTests}). Fourth, it provides semi-empirical generalisations of the scaling laws that bridge between the saltation and bedload regimes (section~\ref{GeneralizationBedload}) and shows how they are affected by modifications of the drag law (section~\ref{GeneralizationDragRarefaction}), which may occur, for example, in highly rarefied atmospheres due to drag rarefaction.

\subsection{Simulation data and scaling laws for saltation} \label{SaltationScalingLaws}
\subsubsection{Cessation threshold} \label{CessationThresholdScaling}
Of the physical parameters affecting the shear velocity at the cessation threshold $u_{\ast t}$, the surface air pressure $P$ varies most strongly with the planetary environment. Furthermore, for a given planetary environment, the grain size $d$ is the most strongly varying relevant physical parameter. To isolate the effect of $P$ on $u_{\ast t}$, we normalise $u_{\ast t}$ in terms of relevant parameters that do neither depend on $P$ nor on $d$, $U_{\ast t}\equiv u_{\ast t}/(\mu g/\rho_p)^{1/3}$ (using that the dynamic viscosity $\mu=\rho_f\nu$ does not depend on $P$), and compare it with the density ratio $s$, which incorporates the effect of $P$ isolated from that of $d$.

\begin{figure}
 \centerline{\includegraphics[width=\textwidth]{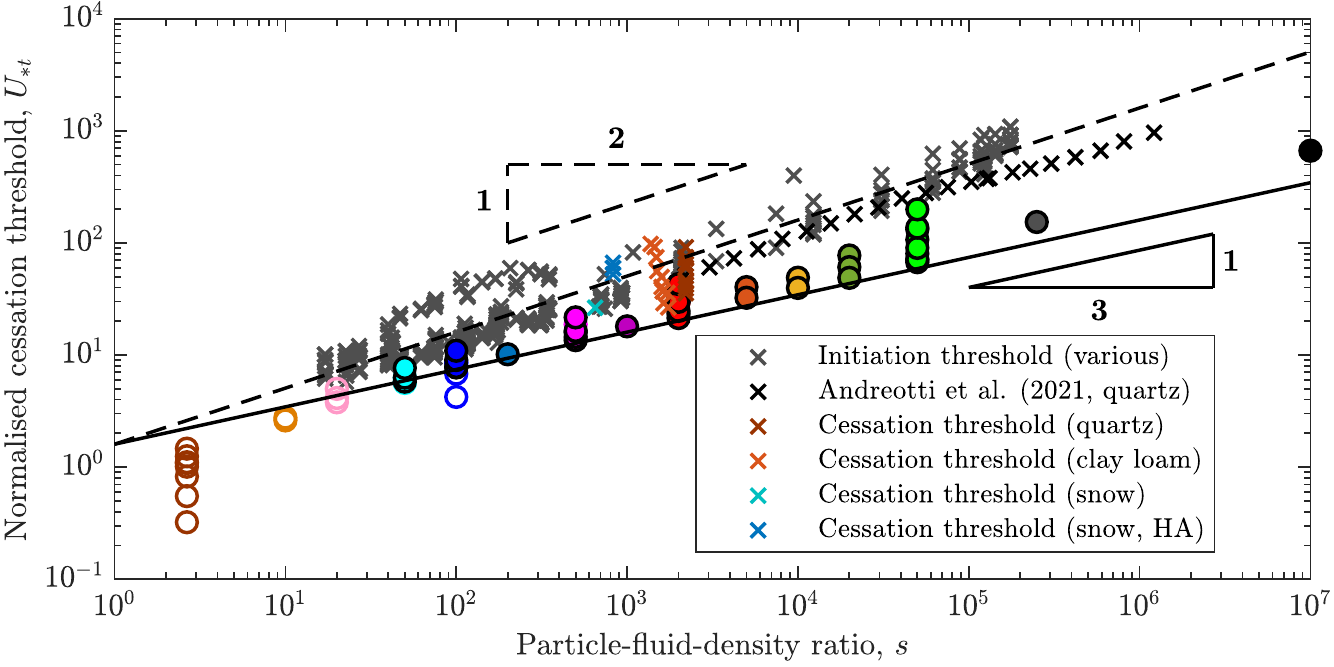}}
\caption{Cessation threshold shear velocity normalised using air-pressure- and grain-size-independent natural units $U_{\ast t}\equiv u_{\ast t}/(\mu g/\rho_p)^{1/3}$ versus density ratio $s$. Symbols that appear in the legend correspond to initiation \citep{Greeleyetal76,Greeleyetal80,IversenWhite82,Greeleyetal84,Burretal15,Burretal20,Swannetal20} and cessation threshold measurements for aeolian transport of quartz \citep{Bagnold37,MartinKok18,Zhuetal19}, clay loam \citep{Chepil45} and snow at sea level \citep{Sugiuraetal98} and high altitude \citep[][HA]{Cliftonetal06}. The dynamic-threshold measurements by \citet{Andreottietal21} may constitute data of the continuous-transport threshold rather than the cessation threshold (discussed in section~\ref{DynamicThresholdMeasurements}). Symbols that do not appear in the legend correspond to numerical simulations for various combinations of $s$ and the Galileo number $Ga$ (see table~\ref{SimulationPhaseSpace} and figure~\ref{PhaseSpace}), with open and filled symbols indicating bedload and saltation conditions, respectively (see figure~\ref{PhaseSpace} for the definition). The solid line corresponds to $U_{\ast t}\propto s^{1/3}$ and represents the lower bound for cessation and initiation thresholds of saltation.}
 \label{LowerBoundThreshold}
\end{figure}

For saltation, the simulations reveal a lower bound for $U_{\ast t}$ scaling as $s^{1/3}$ (figure~\ref{LowerBoundThreshold}, filled circles).
This is distinct from the classical scaling of the saltation initiation threshold with $s^{1/2}$ \citep{Greeleyetal76,Greeleyetal80,IversenWhite82,Greeleyetal84,Burretal15,Burretal20,Swannetal20} (figure~\ref{LowerBoundThreshold}, gray crosses), which follows from a balance between flow-induced and resisting forces or torques acting in bed surface grains \citep{Pahtzetal20a}. Roughly the same $s^{1/2}$-scaling was also found for the dynamic-threshold measurements by \citet{Andreottietal21} carried out in a low-pressure wind tunnel (figure~\ref{LowerBoundThreshold}, black crosses). As mentioned in the introduction and discussed in more detail in section~\ref{DynamicThresholdMeasurements}, these measurements may constitute data of the continuous-transport threshold rather than the cessation threshold.

\begin{figure}
 \centerline{\includegraphics{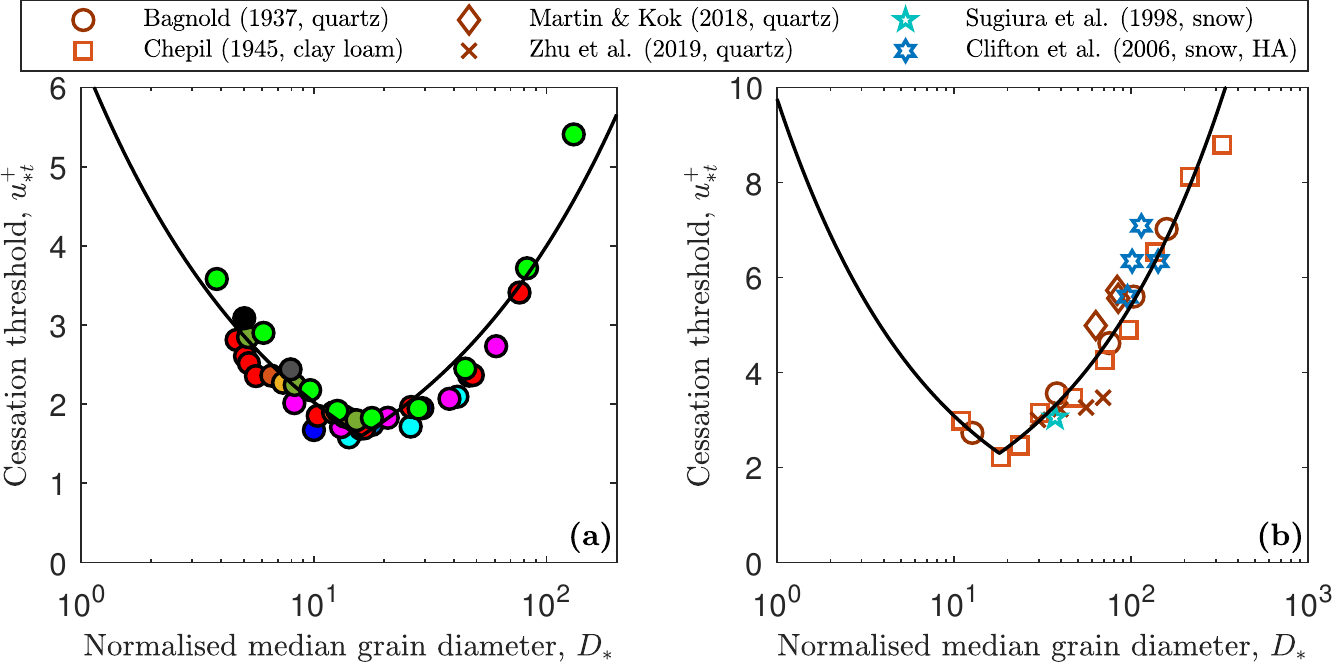}}
\caption{Rescaled cessation threshold shear velocity $u^+_{\ast t}$ versus normalised median grain diameter $D_\ast\equiv\sqrt{s}d^+$. Symbols in (a) correspond to numerical simulations of saltation (see figure~\ref{PhaseSpace} for the definition) for various combinations of the density ratio $s$ and Galileo number $Ga$ (see table~\ref{SimulationPhaseSpace} and figure~\ref{PhaseSpace}). Symbols in (b) correspond to experimental cessation threshold data (see legend) for terrestrial aeolian saltation of quartz \citep{Bagnold37,MartinKok18,Zhuetal19}, clay loam \citep{Chepil45} and snow at sea level \citep{Sugiuraetal98} and high altitude \citep[][HA]{Cliftonetal06}. The solid lines correspond to (\ref{Ut}), with $(D_\ast^{\rm min},u^{+\rm min}_{\ast t})=(16,1.6)$ in (a) and $(D_\ast^{\rm min},u^{+\rm min}_{\ast t})=(18,2.3)$ in (b).}
 \label{ThresholdScaling}
\end{figure}

In addition to its $s^{1/3}$-scaling, $U_{\ast t}$ varies with the normalised median grain diameter $D_\ast\equiv\sqrt{s}d^+=\sqrt{s}d\tilde g/(\tilde g\nu)^{2/3}$, described by the following relationship between the rescaled cessation threshold $u_{\ast t}^+$ (note that $u_{\ast t}^+=U_{\ast t}/(s-1)^{1/3}$) and $D_\ast$:
\begin{equation}
 u^+_{\ast t}=u^{+\rm min}_{\ast t}\max\left[\left(\frac{D_\ast}{D^{\rm min}_\ast}\right)^{-1/2},\left(\frac{D_\ast}{D^{\rm min}_\ast}\right)^{1/2}\right]. \label{Ut}
\end{equation}
It contains the parameters $D^{\rm min}_\ast$ and $u^{+\rm min}_{\ast t}$, which denote the location and magnitude, respectively, of the minimum of the function $u_{\ast t}^+(D_\ast)$, corresponding to the lower bound of $U_{\ast t}$ for saltation in figure~\ref{LowerBoundThreshold}. Equation~(\ref{Ut}) is consistent with the simulations (figure~\ref{ThresholdScaling}(a)) and experiments (figure~\ref{ThresholdScaling}(b)) for the saltation regime, though with slightly different parameter values: $(D^{\rm min}_\ast,u^{+\rm min}_{\ast t})=(16,1.6)$ versus $(D^{\rm min}_\ast,u^{+\rm min}_{\ast t})=(18,2.3)$, respectively.
The associated relative change of $u_{\ast t}^+$ by $2.3/1.6\simeq1.4$ is well within the typical systematic uncertainty of cessation threshold measurements. For example, \citet{Creysselsetal09} reported $\Theta_t=0.009$ for their terrestrial wind tunnel experiments ($d=242~\mathrm{\mu m}$), obtained from extrapolating transport rate measurements to vanishing transport using the transport rate model of \citet{UngarHaff87}, whereas \citet{PahtzDuran20} reported $\Theta_t=0.0035$ for the very same data using a different transport rate model for the extrapolation, resulting in a relative change of $\sqrt{0.009/0.0035}\simeq1.6$.

\subsubsection{Equilibrium transport rate}
\begin{figure}
\centerline{\includegraphics{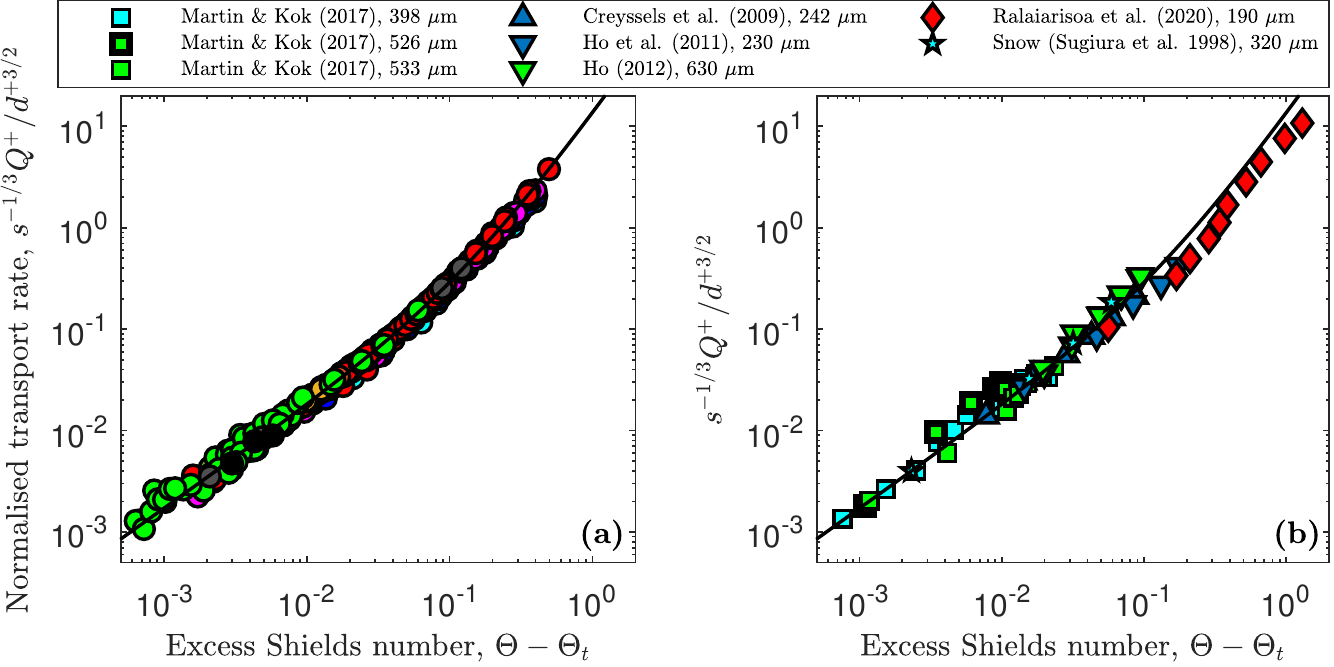}}
\caption{Normalised sediment transport rate $s^{-1/3}Q^+/d^{+3/2}$ versus Shields number in excess of the cessation threshold $\Theta-\Theta_t$. Symbols in (a) correspond to numerical simulations of saltation (see figure~\ref{PhaseSpace} for the definition) for various combinations of the density ratio $s$ and Galileo number $Ga$ (see table~\ref{SimulationPhaseSpace} and figure~\ref{PhaseSpace}) with $Ga\sqrt{s}>81$, and Shields number $\Theta$. Symbols in (b) correspond to measurements for different grain sizes (indicated in the legend) for terrestrial aeolian saltation of minerals \citep{Creysselsetal09,Hoetal11,Ho12,MartinKok17,Ralaiarisoaetal20} and snow \citep{Sugiuraetal98}. The values of $\Theta_t$ in (b) for a given experimental data set are obtained from extrapolating (\ref{Q}) to vanishing transport. Note that \citet{Ralaiarisoaetal20} reported that transport may not have been completely in equilibrium in their experiments. The solid lines correspond to (\ref{Q}).}
 \label{QScalingSaltation}
\end{figure}

The simulations of saltation and experiments reasonably collapse on the master curve (figure~\ref{QScalingSaltation})
\begin{equation}
 Q^+/d^{+3/2}=1.7s^{1/3}(\Theta-\Theta_t)+12s^{1/3}(\Theta-\Theta_t)^2 \label{Q}
\end{equation}
if $Ga\sqrt{s}>81$. The vast majority of planetary transport occurring in nature and most of the simulated saltation conditions satisfy this criterion. Note that $Ga\sqrt{s}$ can be interpreted as a Stokes-like number \citep{Berzietal16}, encoding the importance of grain inertia relative to viscous drag forcing, and controls the transition to viscous bedload \citep{Pahtzetal21}.

\subsection{Physical justifications of saltation scaling laws} \label{PhysicalJustification}
\subsubsection{First-principle-based proof that $u_{\ast t}^+$ is a function of only $D_\ast$} \label{Proof}
In general, the shear velocity at the cessation threshold $u_{\ast t}$ is a function of the five control parameters $\rho_p$, $\rho_f$, $\nu$, $\tilde g$ and $d$ \citep{ClaudinAndreotti06}. These parameters involve three units (mass, length and time). According to the $\Pi$ theorem \citep{Barenblatt96}, the physical system, and therefore any dimensionless system property such as $u_{\ast t}^+$, is then controlled by two dimensionless numbers, for example the density ratio $s$ and the normalised median grain diameter $D_\ast$:
\begin{equation}
 u^+_{\ast t}=f(s,D_\ast). \label{ut_vsgeneral}
\end{equation}
To determine the function $f$ in (\ref{ut_vsgeneral}), existing cessation threshold models have made various idealisations of the fluid-particle system \citep{ClaudinAndreotti06,Kok10b,Berzietal16,Berzietal17,PahtzDuran18a,Andreottietal21,Pahtzetal21,GunnJerolmack22}. In particular, they all drastically coarse-grain the particle phase of the aeolian transport layer above the bed surface, either by representing the entire grain motion by identical periodic saltation trajectories \citep{ClaudinAndreotti06,Kok10b,Berzietal16,Berzietal17,PahtzDuran18a,Andreottietal21,Pahtzetal21,GunnJerolmack22} or by an average motion behaviour \citep{Kok10b,PahtzDuran18a} that is mathematically equivalent to an identical periodic trajectory representation \citep{Pahtzetal20a}.

Here, in contrast to previous models, we do not resort to any such coarse-graining. Instead, we idealise the system in the following comparably mild manner.
\begin{enumerate}
 \item We consider only buoyancy and Stokes drag as fluid--grain interactions, neglecting form drag contributions. This would be justified if relatively fast saltating grains dominated the near-threshold grain dynamics, since comparably faster grains exhibit comparably lower fluid-particle velocity differences and, thus, comparably less form drag relative to Stokes drag.
 \item Due to the typically relatively small shear Reynolds numbers associated with planetary transport near the cessation threshold, $Ga\sqrt{\Theta_t}\lesssim10$, we consider a smooth inner turbulent boundary layer mean flow velocity profile $u_x(z)$, neglecting hydrodynamically rough contributions (and turbulent fluctuations, which are also neglected in the numerical simulations).
 \item Since vanishingly few grains are in motion sufficiently close to the cessation threshold, we neglect the feedback of the grain motion on the flow.
 \item Since saltation trajectories are typically much larger than the grain size, we consider an idealised flat bed and assume that the zero level of the flow velocity coincides with the grain elevation at grain--bed impact ($z=0$), neglecting the effect of the flow very near the bed surface to the overall grain motion.
 \item While we do not specify the distribution of grain lift-off velocities $f_\uparrow$ and grain impact velocities $f_\downarrow$, we assume that the boundary conditions mapping $f_\downarrow$ to $f_\uparrow$ in the steady state are scale-free, as for grain--bed rebounds \citep{Beladjineetal07}, neglecting the potential effect of $\sqrt{\tilde gd}$ on grain--bed collisions. Most grains ejected by the splash of a grain impacting the bed with velocity $\boldsymbol{v}_\downarrow$ exhibit a velocity on the order of $\sqrt{\tilde gd}$ and only the few grains corresponding to the upper-tail end of the distribution exhibit an ejection velocity proportional to $|\boldsymbol{v}_\downarrow|$ \citep{Lammeletal17}. Hence, this assumption effectively means that grain--bed rebounds and/or rare extreme ejection events dominate the saltation dynamics relevant for the cessation threshold scaling.
\end{enumerate}

Under the above assumptions, the equations of motion for a given grain are \citep{Pahtzetal21}
\begin{align}
 \dot v^+_z&=-1-v^+_z/v^+_s, \label{vz} \\
 \dot v^+_x&=(u^+_x-v^+_x)/v^+_s, \label{vx} \\
 u^+_x&=u^+_\ast f_u(u^+_\ast z^+), \label{ux}
\end{align}
where $\boldsymbol{v}^+$ is the rescaled grain velocity, $v_s^+=4sd^{+2}/(3\Rey_c)$ the rescaled Stokes settling velocity (obtained from the high-viscosity limit of (\ref{Drag})), and $f_u(X)$ denotes a function describing $u_x/u_\ast$ for an undisturbed smooth inner turbulent boundary layer. It obeys $f_u(X)=X$ within the viscous sublayer of the turbulent boundary layer ($X\lesssim5$) and $f_u(X)\simeq\kappa^{-1}\ln(9X)$ within its log-layer ($X\gtrsim30$). Extrapolated into the transitional buffer layer in between, both profiles would intersect at about $X=11$, which is why $\delta_\nu=11\nu/u_\ast$ is termed viscous-sublayer thickness.

Parametrised by $v^+_s$ and $u^+_\ast$, (\ref{vz})-(\ref{ux}) map $f_\uparrow$ to $f_\downarrow$. Combined with the scale-free boundary conditions, mapping $f_\downarrow$ back to $f_\uparrow$, they imply that the grain motion is fully determined by $v^+_s$ and $u^+_\ast$. For a given $v^+_s$, the cessation threshold $u^+_{\ast t}$ then corresponds to the smallest value of $u^+_\ast$ for which a solution of the combined system exists \citep{Pahtzetal21}. This implies that there is a function $f$ mapping $D_\ast=\sqrt{18v^+_s}$ (valid for $\Rey_c=24$, the standard case of non-rarefied drag) to $u^+_{\ast t}$:
\begin{equation}
 u^+_{\ast t}=f(D_\ast). \label{ut_vs}
\end{equation}
In summary, the above assumptions simplify the general two-parametric dependence of $u_{\ast t}^+$ in (\ref{ut_vsgeneral}) to the one-parametric dependence in (\ref{ut_vs}), in agreement with (\ref{Ut}).

\subsubsection{Semi-empirical model of cessation threshold scaling}
\begin{figure}
 \centerline{\includegraphics{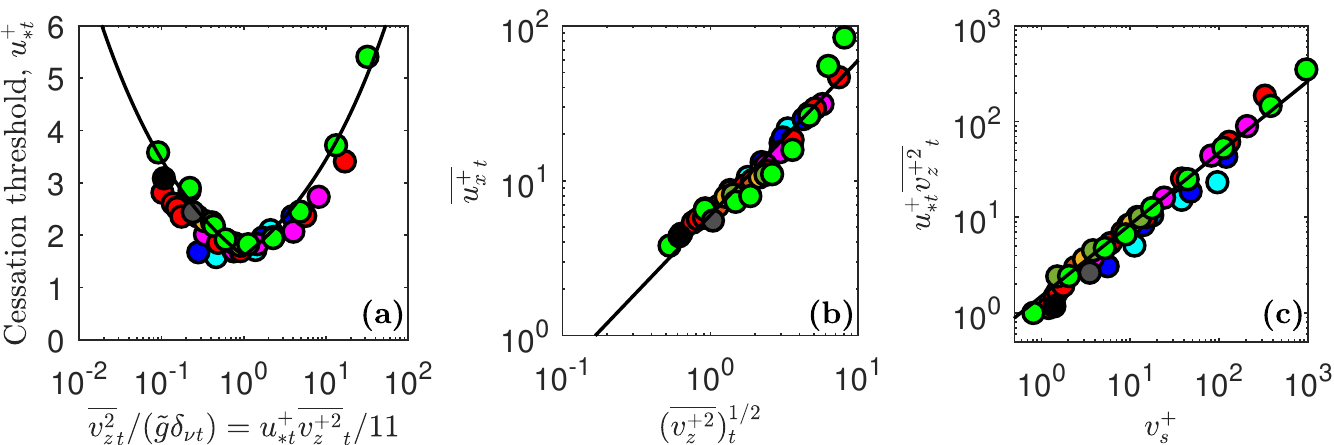}}
\caption{(a) Rescaled threshold shear velocity $u^+_{\ast t}$ versus ratio between hop height $\overline{v_z^2}_t/\tilde g$ and viscous-sublayer thickness $\delta_{\nu t}=11\nu/u_{\ast t}$ near the cessation threshold. (b) Rescaled transport layer-averaged fluid velocity $\overline{u^+_x}_t$ versus $(\overline{v_z^{+2}})^{1/2}_t$ near the cessation threshold. (c) Plot of $u^+_{\ast t}\overline{v_z^{+2}}_t$ versus $v^+_s$. Symbols correspond to numerical simulations of saltation (see figure~\ref{PhaseSpace} for the definition) for various combinations of the density ratio $s$ and Galileo number $Ga$ (see table~\ref{SimulationPhaseSpace} and figure~\ref{PhaseSpace}). The solid lines in (a), (c) and (b) correspond to (\ref{ut_delta}), (\ref{delta_vs}) and $\overline{u^+_x}_t=6(\overline{v_z^{+2}})^{1/2}_t$, respectively.}
 \label{LowerBoundThreshold2}
\end{figure}

While the above analysis explains why $u_{\ast t}^+=f(D_\ast)$ in (\ref{Ut}), it does not yield the function $f$ itself. Here, we derive the expression for $f$ in (\ref{Ut}) guided by the simulations. The latter show that the minimum $u^{+\rm min}_{\ast t}$ for saltation occurs when the hop height $\overline{v_z^2}_t/\tilde g$ is about equal to the viscous-sublayer thickness $\delta_{\nu t}=11\nu/u_{\ast t}$ near the cessation threshold (figure~\ref{LowerBoundThreshold2}(a)).
This can be explained using the empirical, yet physically reasonable, simulation-supported proportionality between the average fluid velocity $\overline{u_x}_t$ and $(\overline{v_z^2})^{1/2}_t$ near the cessation threshold (figure~\ref{LowerBoundThreshold2}(b)). In fact, averaging (\ref{ux}) over all grain trajectories and the transport layer, using the approximation $\overline{f_u(u^+_{\ast t}z^+)}_t\simeq f_u(u^+_{\ast t}\overline{z^+}_t)$, and using this proportionality approximately yields for saltation ($\overline{z^+}_t\simeq\overline{v_z^{+2}}_t$, see figure~\ref{PhaseSpace}):
\begin{equation}
 u^+_{\ast t}\propto\left[\frac{u^+_{\ast t}\overline{v_z^{+2}}_t}{f_u^2\left(u^+_{\ast t}\overline{v_z^{+2}}_t\right)}\right]^{1/3}.
\end{equation}
Within the viscous sublayer ($u^+_{\ast t}\overline{v_z^{+2}}_t\lesssim5$), this relation simplifies to $u^+_{\ast t}\propto(u^+_{\ast t}\overline{v_z^{+2}}_t)^{-1/3}$ and within the log-layer approximately to $u^+_{\ast t}\propto(u^+_{\ast t}\overline{v_z^{+2}}_t)^{1/3}$, neglecting the logarithmic term. The crossover between the two power laws occurs about at $u^+_{\ast t}\overline{v_z^{+2}}_t=11$, that is, when the hop height exceeds the viscous-sublayer thickness ($\overline{v_z^2}_t/\tilde g=\delta_{\nu t}$). Hence, the parabolic law
\begin{equation}
 u^+_{\ast t}=u^{+\rm min}_{\ast t}\max\left[\left(\frac{\overline{v_z^2}_t}{\tilde g\delta_{\nu t}}\right)^{-1/3},\left(\frac{\overline{v_z^2}_t}{\tilde g\delta_{\nu t}}\right)^{1/3}\right] \label{ut_delta}
\end{equation}
fits the saltation data reasonably well (solid line in figure~\ref{LowerBoundThreshold2}(a)).

Following from the analysis we have used to deduce (\ref{ut_vs}), the grain kinematics near the cessation threshold, and thus $\overline{v_z^{+2}}_t$, should be controlled by $u^+_{\ast t}$ or $v^+_s$. Indeed, the simulations of saltation suggest the empirical relation (figure~\ref{LowerBoundThreshold2}(c))
\begin{equation}
 u^+_{\ast t}\overline{v_z^{+2}}_t=1.5v_s^{+3/4}, \label{delta_vs}
\end{equation}
which leads to (\ref{Ut}) with $D_\ast^{\rm min}=\sqrt{18}(11/1.5)^{2/3}\simeq16$.

According to the above model, the grain size scaling of $u^+_{\ast t}$ in (\ref{Ut}), despite being mathematically equivalent to the well-known cohesive ($u_{\ast t}\sim d^{-1/2}$, left branch) and cohesionless ($u_{\ast t}\sim d^{1/2}$, right branch) limits of the saltation initiation threshold \citep{ShaoLu00}, follows purely from hydrodynamics rather than the onset of cohesion at small grain sizes.

\subsubsection{Physics behind equilibrium transport rate scaling} \label{PhysicsTransportRate}
Analytical, physical models of the equilibrium transport rate $Q$ for aeolian saltation typically separate it into the mass of transported sediment per unit area of the bed $M$ and its average streamwise velocity $V$ through $Q=MV$. In most models, it is reasoned that the scaling of $V$ is in one way or another linked to grain--bed collisions, and since the average outcome of grain--bed collisions should be roughly independent of the wind speed at equilibrium, $V$ is taken as equal to its near-threshold value $V_t$ \citep{UngarHaff87,Duranetal11,Koketal12,Berzietal16}. However, it has been shown that, for sufficiently intense saltation, midair collisions significantly disturb grain trajectories \citep{Carneiroetal13,PahtzDuran20,Ralaiarisoaetal20}, leading to an additional additive term increasing as $M^+/d^+$ \citep{PahtzDuran20}:
\begin{equation}
 Q^+=M^+V^+_t(1+c_MM^+/d^+), \label{QM}
\end{equation}
where $c_M$ is a constant parameter. It is not trivial to evaluate the scalings of $M^+$ and $V^+_t$ with the simulation data, since extracting $M$ and $V$ from DEM-based numerical transport simulations is ambiguous \citep{Duranetal12,PahtzDuran18b}. One possible way is to define $M$ as the mass $M_0$ of grains moving above the bed surface ($z=0$) per unit bed area and $V$ as their average streamwise velocity \citep{PahtzDuran18b}: 
\begin{align}
 M&\equiv\int_0^\infty\rho\d z=M_0, \label{M0} \\
 V&\equiv\frac{\int_0^\infty\rho\langle v_x\rangle\d z}{\int_0^\infty\rho\d z}=\overline{v_x}. \label{vx0}
\end{align}
This definition uses that most (but not all) sediment transport occurs at elevations $z>0$, especially for saltation and, therefore, $M_0\overline{v_x}=\int_0^\infty\rho\langle v_x\rangle\d z\simeq\int_{-\infty}^\infty\rho\langle v_x\rangle\d z=Q$ \citep{PahtzDuran18b}. Alternatively, one can define $V$ as the mass flux-weighted average $\overline{v_x}^q$ of the streamwise velocity of all grains and $M_q$, the associated value of $M$, as $M_q\equiv Q/\overline{v_x}^q$ \citep{Duranetal12}:
\begin{align}
 M&\equiv\frac{\left(\int_{-\infty}^\infty\rho\langle v_x\rangle\d z\right)^2}{\int_{-\infty}^\infty\rho\langle v_x^2\rangle\d z}=M_q, \label{Mq} \\
 V&\equiv\frac{\int_{-\infty}^\infty\rho\langle v_x^2\rangle\d z}{\int_{-\infty}^\infty\rho\langle v_x\rangle\d z}=\overline{v_x}^q, \label{vxq}
\end{align}
where $\overline{\cdot}^q\equiv\frac{1}{Q}\int_{-\infty}^\infty\rho\langle v_x\cdot\rangle\d z$.

\begin{figure}
 \centerline{\includegraphics{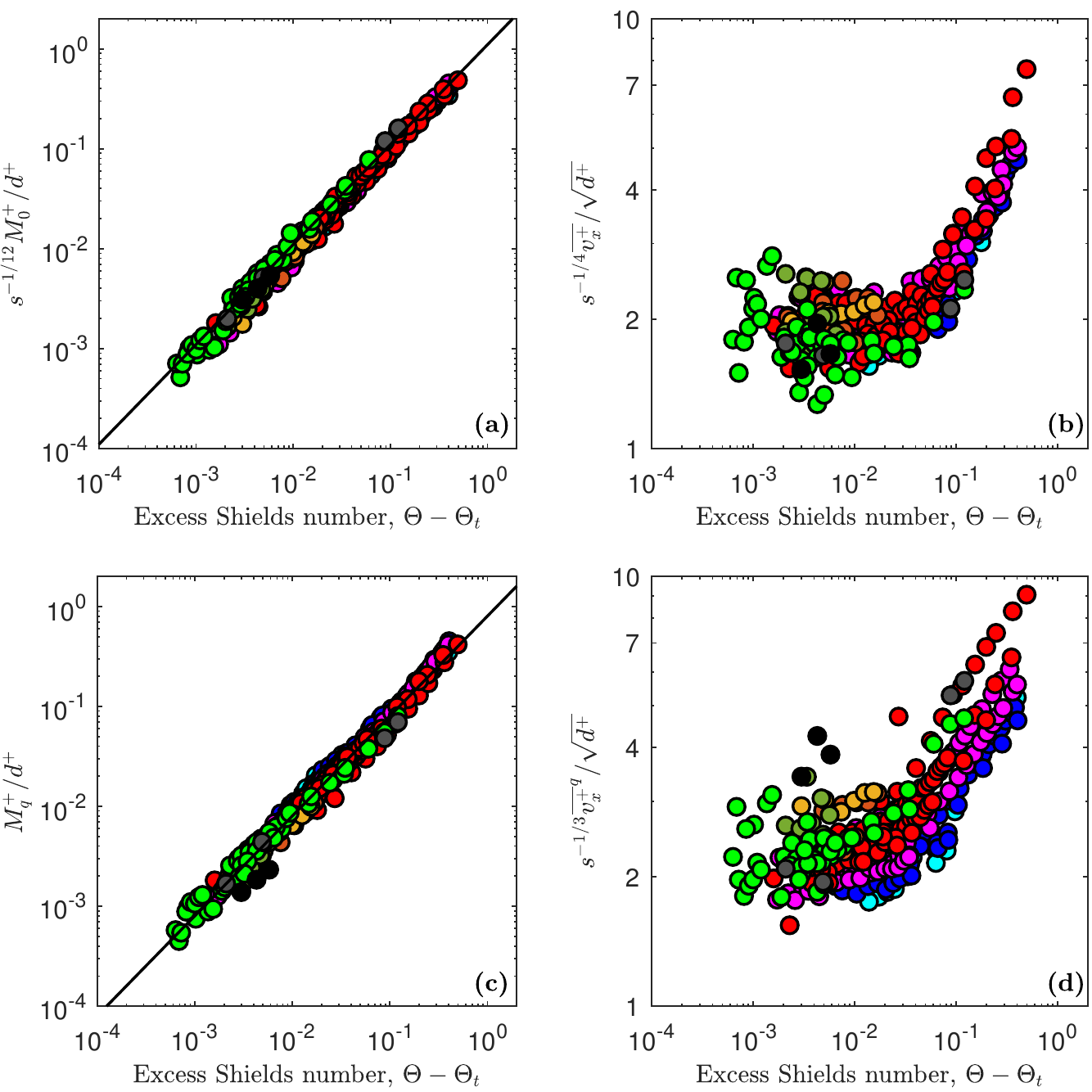}}
\caption{(a) and (c) Normalised transport loads $s^{-1/12}M^+_0/d^+$ and $M^+_q/d^+$, using the definitions (\ref{M0}) and (\ref{Mq}), respectively, of $M$; and (b) and (d) normalised average streamwise grain velocities $s^{-1/4}\overline{v^+_x}/\sqrt{d^+}$ and $s^{-1/3}\overline{v^+_x}^q/\sqrt{d^+}$, using the definition (\ref{vx0}) and (\ref{vxq}), respectively, of $V$ versus Shields number in excess of the cessation threshold $\Theta-\Theta_t$. Symbols correspond to numerical simulations of saltation (see figure~\ref{PhaseSpace} for the definition) for various combinations of the density ratio $s$ and Galileo number $Ga$ (see table~\ref{SimulationPhaseSpace} and figure~\ref{PhaseSpace}) with $Ga\sqrt{s}>81$, and Shields number $\Theta$. The solid lines in (a) and (b) correspond to the left equations in (\ref{M0Scaling}) and (\ref{MqScaling}), respectively.}
 \label{MVSeparation}
\end{figure}

For the above two definitions of $M$ and $V$, the simulations are roughly described by scaling laws in which a comparably small part of the $s^{1/3}$-scaling factor in (\ref{Q}) goes into $M^+/d^+$ and a comparably large part into $V^+_t/\sqrt{d^+}$ (figure~\ref{MVSeparation}).
However, the exact partitioning of $s^{1/3}$ depends on the chosen definition (figures~\ref{MVSeparation}(a) and \ref{MVSeparation}(b) versus figures~\ref{MVSeparation}(c) and \ref{MVSeparation}(d)):
\begin{align}
 M^+_0&\propto s^{1/12}d^+(\Theta-\Theta_t),&\overline{v^+_x}_t&\propto s^{1/4}\sqrt{d^+}, \label{M0Scaling} \\
 M^+_q&\propto d^+(\Theta-\Theta_t),&\overline{v^+_x}^q_t&\propto s^{1/3}\sqrt{d^+}. \label{MqScaling}
\end{align}
The latter scaling is consistent with the prediction $M^+\propto d^+(\Theta-\Theta_t)$ from physical models \citep{UngarHaff87,Duranetal11,Berzietal16,PahtzDuran20} and with (\ref{Q}) when combined with (\ref{QM}). However, it means that $V^+_t\propto s^{1/3}\sqrt{d^+}$, which is a highly unusual scaling, different from the existing models $V^+_t\propto\sqrt{d^+}$ \citep{UngarHaff87,Berzietal16} and $V^+_t\propto u^+_{\ast t}$ \citep{Duranetal11,Koketal12,PahtzDuran20}.

\subsection{Test of existing models against simulations of saltation} \label{ModelTests}
\subsubsection{Test of cessation threshold models} \label{CessationThresholdModelTest}
\begin{figure}
 \centerline{\includegraphics{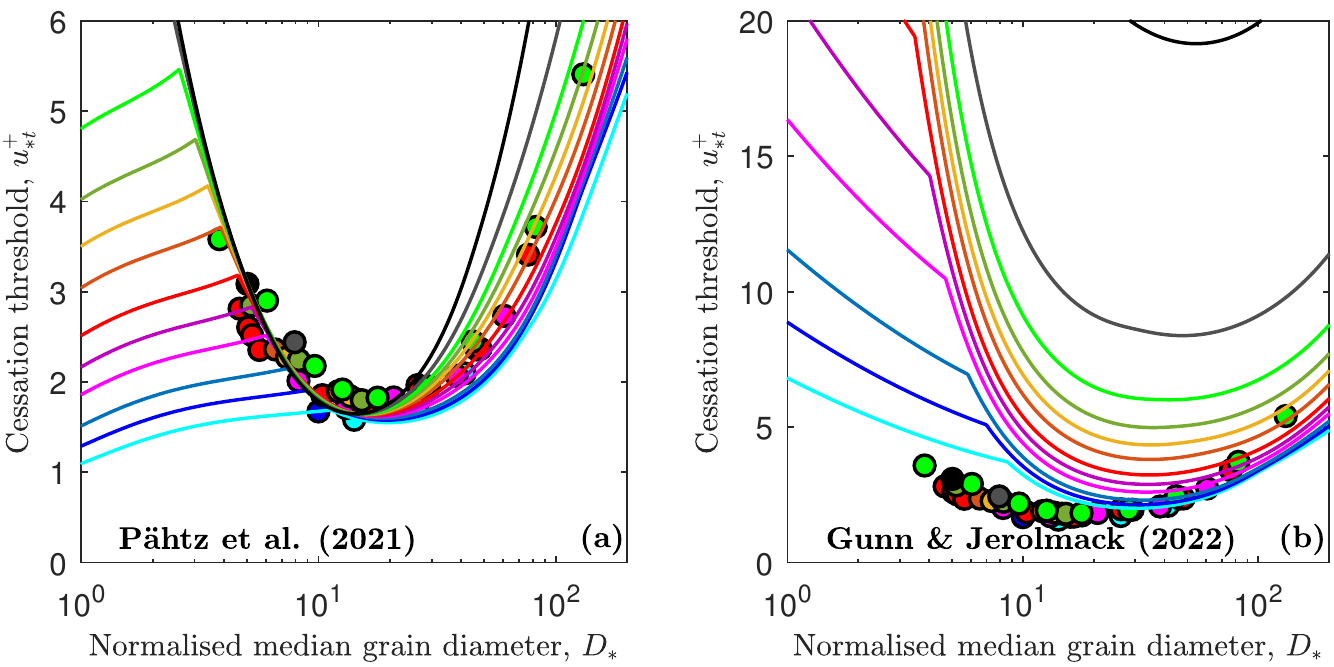}}
\caption{Evaluation of the cessation threshold models of (a) \citet{Pahtzetal21} and (b) \citet{GunnJerolmack22}, where the latter's drag and lift laws have been modified to those employed in the simulations for a fair comparison. Rescaled cessation threshold $u_{\ast t}^+$ versus normalised median grain diameter $D_\ast\equiv\sqrt{s}d^+$. Symbols correspond to numerical simulations of saltation (see figure~\ref{PhaseSpace} for the definition) for various combinations of the density ratio $s$ and Galileo number $Ga$ (see table~\ref{SimulationPhaseSpace} and figure~\ref{PhaseSpace}). The solid lines indicate the respective model predictions. Their color characterises $s$ in accordance with the symbol color.}
 \label{CessationThresholdModelComparison}
\end{figure}

The most important assumption that led to the simulation-supported statement that the rescaled cessation threshold $u_{\ast t}^+$ is solely controlled by the normalised median grain diameter $D_\ast$ in section~\ref{Proof} is that of scale-free boundary conditions. The only existing cessation threshold model with scale-free boundary condition is that of \citet{Pahtzetal21}, which we here compare with the most recent alternative, that of \citet{GunnJerolmack22}. The latter's most important feature is that it superimposes a $Ga$-dependent damping on the scale-free laws describing grain--bed rebounds, where the damping function is essentially fitted to agreement with experimental cessation threshold data. We find that, while the model of \citet{Pahtzetal21} captures the simulation data very well, the model of \citet{GunnJerolmack22}, with its drag and lift laws being modified to those employed in the simulations (i.e., (\ref{Drag}) and no lift) for a fair comparison, is in very strong disagreement (figure~\ref{CessationThresholdModelComparison}).
This is discussed in section~\ref{DynamicThresholdMeasurements}.

\subsubsection{Test of equilibrium transport rate models}
\begin{figure}
 \centerline{\includegraphics{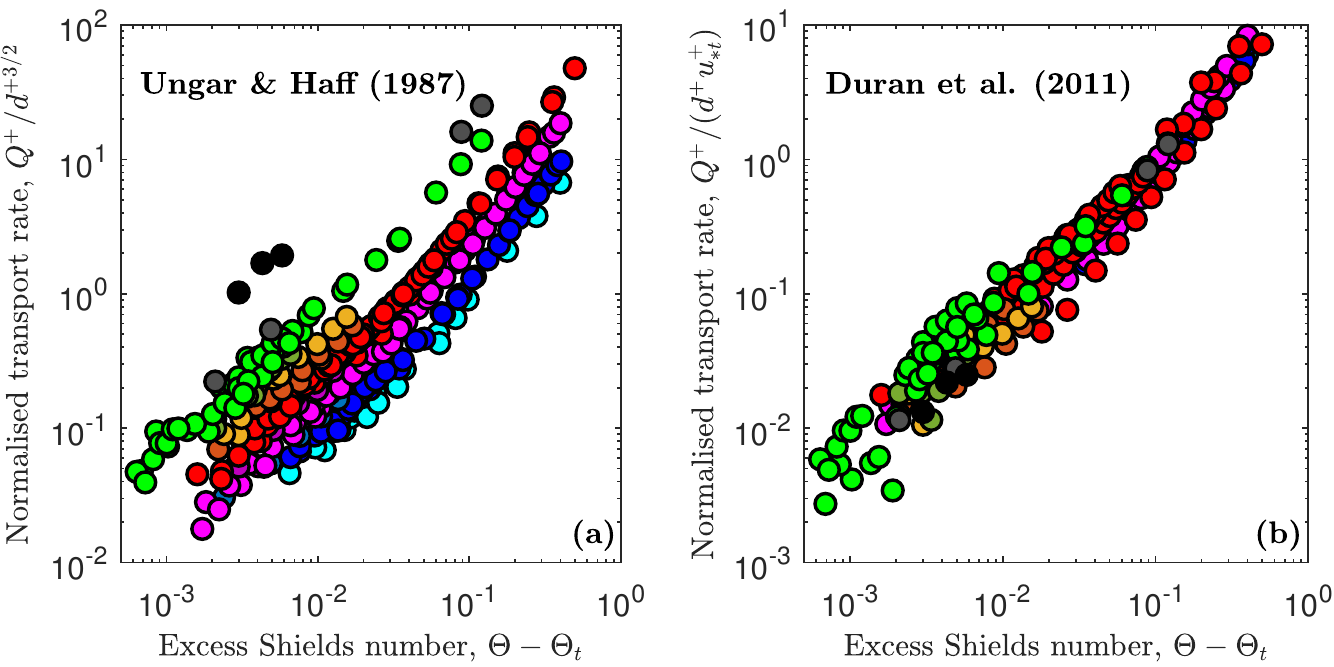}}
\caption{Evaluation of the physically based functional relationships for the sediment transport rate by \citet{UngarHaff87} and \citet{Duranetal11}. Normalised sediment transport rate (a) $Q^+/d^{+3/2}$ and (b) $Q^+/(d^+u^+_{\ast t})$ versus Shields number in excess of the cessation threshold $\Theta-\Theta_t$. Symbols correspond to numerical simulations of saltation (see figure~\ref{PhaseSpace} for the definition) for various combinations of the density ratio $s$ and Galileo number $Ga$ (see table~\ref{SimulationPhaseSpace} and figure~\ref{PhaseSpace}) with $Ga\sqrt{s}>81$, and Shields number $\Theta$.}
 \label{QScalingPrevious}
\end{figure}

The simulations of saltation are not or not well captured by the two most widely used physical models of the equilibrium aeolian transport rate: the model of \citet{UngarHaff87} and others \citep[e.g.,][]{JenkinsValance14,Berzietal16}, $Q^+/d^{+3/2}=f_1(\Theta-\Theta_t)$ (figure~\ref{QScalingPrevious}(a)) and the model of \citet{Duranetal11} and others \citep{Koketal12,PahtzDuran20}, $Q^+/(d^+u^+_{\ast t})=f_2(\Theta-\Theta_t)$ (figure~\ref{QScalingPrevious}(b)).

\subsection{Generalised scaling laws across saltation and turbulent bedload} \label{GeneralizationBedload}
\begin{figure}
 \centerline{\includegraphics{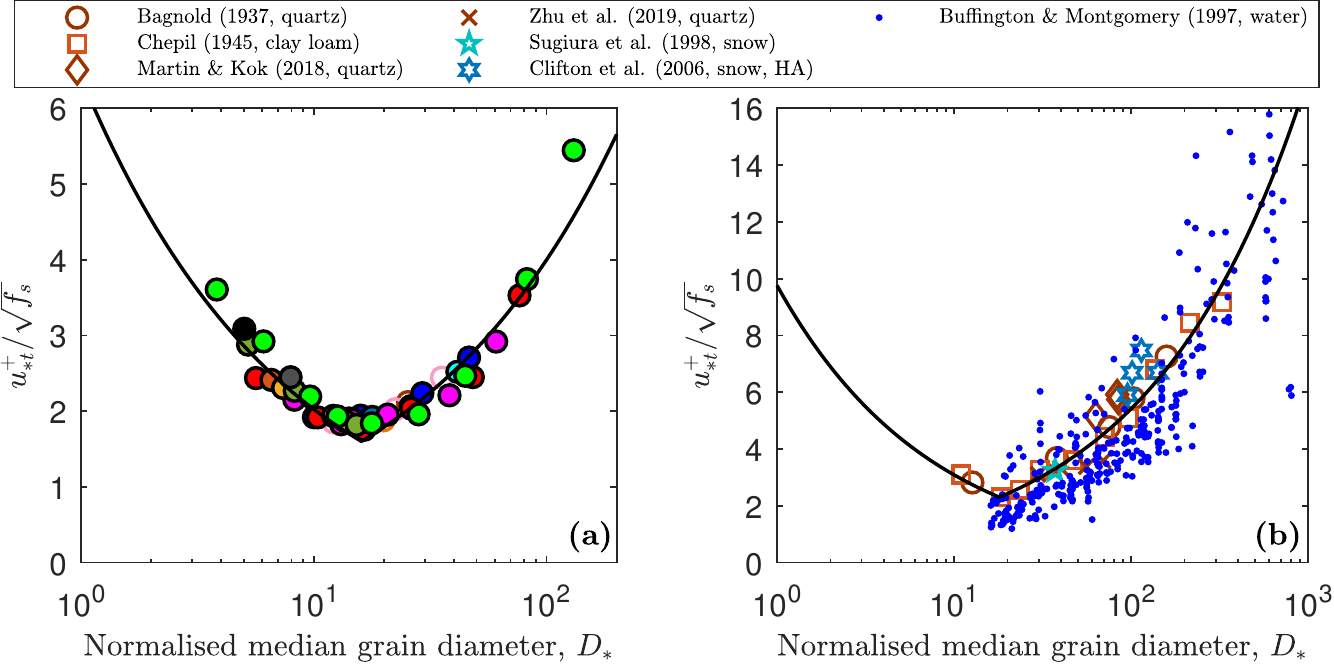}}
\caption{Bedload-corrected rescaled cessation threshold shear velocity $u^+_{\ast t}/\sqrt{f_s}$ versus normalised median grain diameter $D_\ast\equiv\sqrt{s}d^+$. Symbols in (a) correspond to numerical simulations for various combinations of the density ratio $s$ and Galileo number $Ga$ (see table~\ref{SimulationPhaseSpace} and figure~\ref{PhaseSpace}) with $Ga\sqrt{s}>81$, where open and filled symbols indicate bedload and saltation conditions, respectively (see figure~\ref{PhaseSpace} for the definition). Symbols in (b) correspond to experimental cessation threshold data (see the legend) for terrestrial aeolian saltation of quartz \citep{Bagnold37,MartinKok18,Zhuetal19}, clay loam \citep{Chepil45} and snow at sea level \citep{Sugiuraetal98} and high altitude \citep[][HA]{Cliftonetal06}, and a compilation of experimental threshold data for subaqueous bedload \citep{BuffingtonMontgomery97}. Only data with $Ga\sqrt{s}>81$ are shown. The solid lines correspond to (\ref{Ut}), with $(D_\ast^{\rm min},u^{+\rm min}_{\ast t})=(16,1.6)$ in (a) and $(D_\ast^{\rm min},u^{+\rm min}_{\ast t})=(18,2.3)$ in (b).}
 \label{ThresholdScaling2}
\end{figure}

It is possible to semi-empirically generalise (\ref{Ut}) to also include turbulent bedload conditions, defined by $s\lesssim10$ and $D_\ast\gtrsim D_\ast^{\rm min}$ \citep[equivalent to $Gas^{1/4}\gtrsim64$, which ensures that transported grains significantly penetrate the log-layer;][]{PahtzDuran20}. Turbulent bedload not only includes hopping grains but also rolling grains. The threshold shear velocity required to sustain a pure, very slow rolling motion along the bed surface scales as $u_{\ast t}\propto\sqrt{s\tilde gd}$ \citep{Pahtzetal21}, which corresponds to $u^+_{\ast t}\propto s^{1/4}$ at the cessation threshold minimum $D_\ast=D_\ast^{\rm min}$. We find that the empirical relation $u^+_{\ast t}=\sqrt{f_s}u^{+\rm min}_{\ast t}$, with $f_s\equiv(1+\sqrt{10/s})^{-1}$, captures the transition from $u^+_{\ast t}\propto s^{1/4}$ for $s\ll10$ to $u^+_{\ast t}=u^{+\rm min}_{\ast t}$ for $s\gg10$ at $D_\ast=D_\ast^{\rm min}$. The resulting generalised cessation threshold equation is
\begin{equation}
 u^+_{\ast t}=\sqrt{f_s}u^{+\rm min}_{\ast t}\max\left[\left(\frac{D_\ast}{D^{\rm min}_\ast}\right)^{-1/2},\left(\frac{D_\ast}{D^{\rm min}_\ast}\right)^{1/2}\right]. \label{Utgeneral}
\end{equation}
It is consistent with the simulations and experiments across aeolian and fluvial transport conditions with $Ga\sqrt{s}\gtrsim81$ (figure~\ref{ThresholdScaling2}).

\begin{figure}
 \centerline{\includegraphics{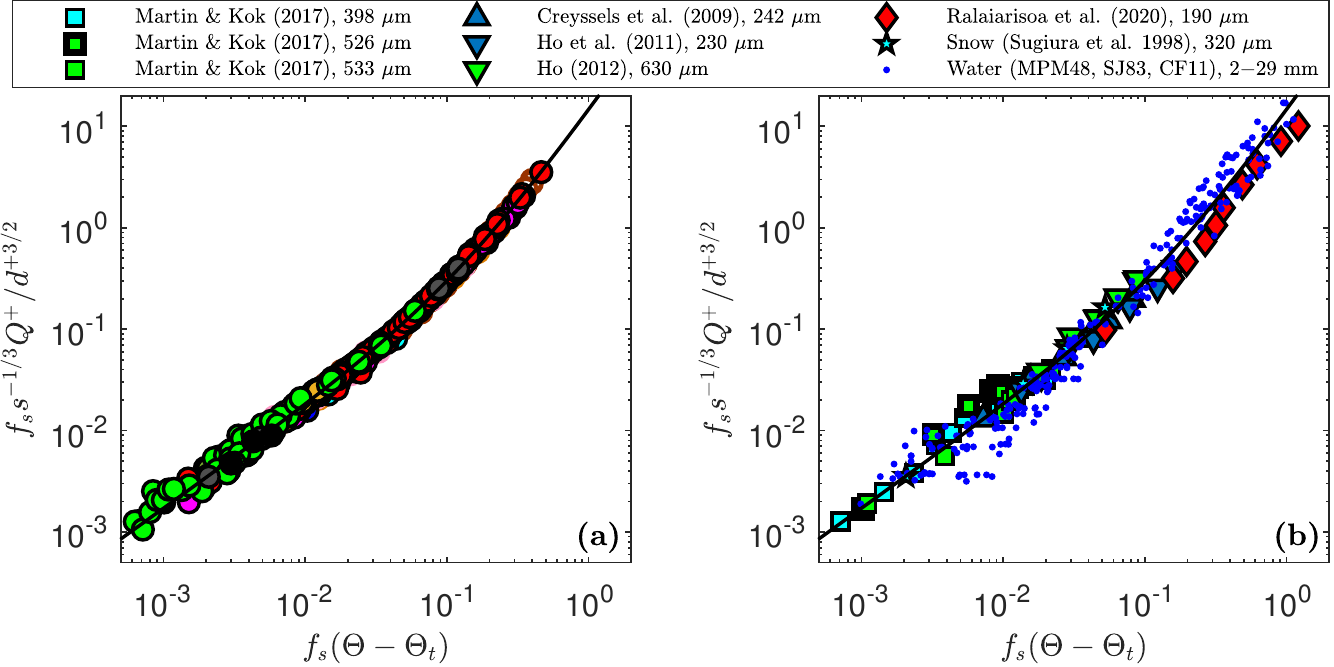}}
\caption{Bedload-corrected normalised sediment transport rate $f_ss^{-1/3}Q^+/d^{+3/2}$ versus bedload-corrected Shields number in excess of the cessation threshold $f_s(\Theta-\Theta_t)$. Symbols in (a) correspond to numerical simulations for various combinations of the density ratio $s$ and Galileo number $Ga$ (see table~\ref{SimulationPhaseSpace} and figure~\ref{PhaseSpace}) with $Ga\sqrt{s}>81$, and Shields number $\Theta$, where open and filled symbols indicate bedload and saltation conditions, respectively (see figure~\ref{PhaseSpace} for the definition). Symbols in (b) correspond to measurements for different grain sizes (indicated in the legend) for terrestrial aeolian saltation of minerals \citep{Creysselsetal09,Hoetal11,Ho12,MartinKok17,Ralaiarisoaetal20} and snow \citep{Sugiuraetal98}, and subaqueous bedload \citep{MeyerPeterMuller48,SmartJaeggi83,CapartFraccarollo11}. We corrected the raw laboratory data by \citet{SmartJaeggi83} and \citet{CapartFraccarollo11} for sidewall drag using the method described in section~2.3 of \citet{Guo15} and for steep bed slopes using $u_\ast^2|_{\rm corrected}=u_\ast^2/f_\alpha$, with $f_\alpha\equiv1-\tan\alpha/0.63$ \citep{Pahtzetal21}. The values of $\Theta_t$ in (b) for a given experimental data set are obtained from extrapolating (\ref{Qgeneral}) to vanishing transport. Note that \citet{Ralaiarisoaetal20} reported that transport may not have been completely in equilibrium in their experiments. The solid lines correspond to (\ref{Qgeneral}).}
 \label{QScalingBedload}
\end{figure}

Furthermore, an empirical generalisation of (\ref{Q}) to turbulent bedload conditions is given by
\begin{equation}
 Q^+/d^{+3/2}=1.7s^{1/3}(\Theta-\Theta_t)+13f_ss^{1/3}(\Theta-\Theta_t)^2, \label{Qgeneral}
\end{equation}
consistent with the simulations and experiments across aeolian and fluvial transport conditions with $Ga\sqrt{s}\gtrsim81$ (figure~\ref{QScalingBedload}).

\begin{figure}
 \centerline{\includegraphics{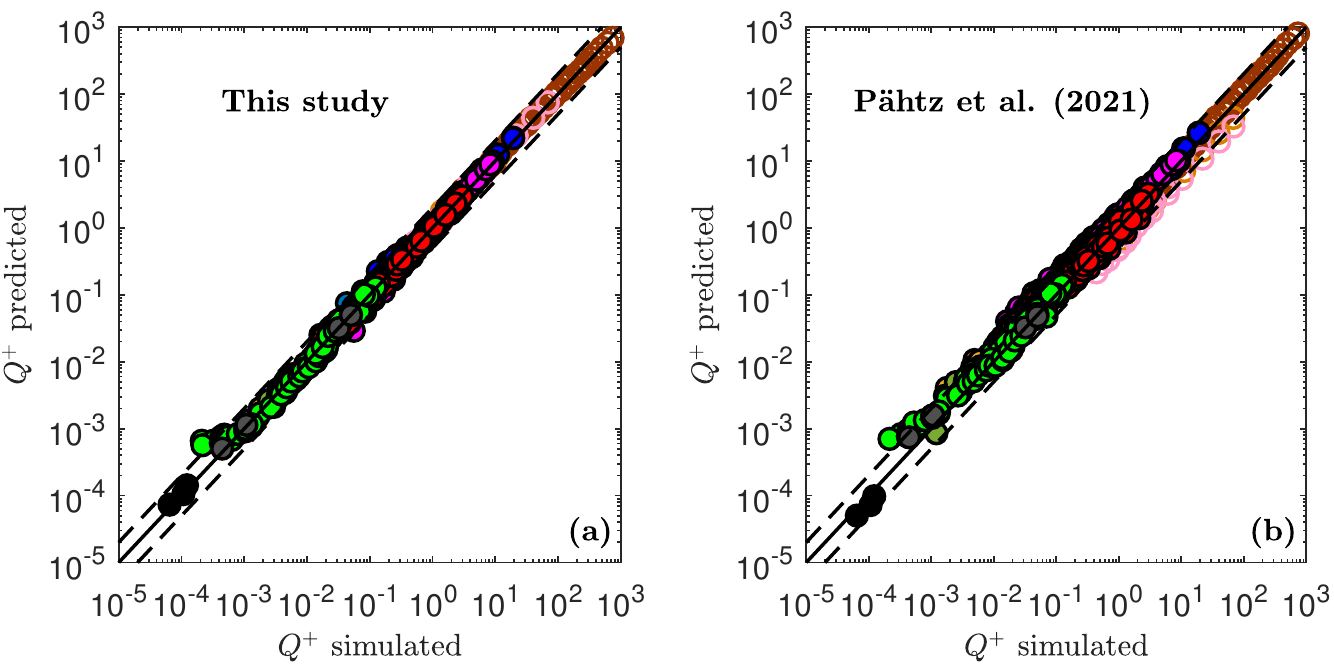}}
 \caption{Predicted versus simulated sediment transport rate $Q^+$. (a) Predictions by (\ref{Ut}) with $(D^{\rm min}_\ast,u^{+\rm min}_{\ast t})=(16,1.6)$ and (\ref{Qgeneral}). (b) Predictions by the model of \citet{Pahtzetal21}. Symbols correspond to numerical simulations for various combinations of the density ratio $s$ and Galileo number $Ga$ (see table~\ref{SimulationPhaseSpace} and figure~\ref{PhaseSpace}) with $Ga\sqrt{s}>81$, where open and filled symbols indicate bedload and saltation conditions, respectively (see figure~\ref{PhaseSpace} for the definition). The solid lines indicate perfect agreement. The dashed lines indicate a deviation by a factor of two.}
 \label{PredictedMeasured}
\end{figure}

Put together, (\ref{Utgeneral}) and (\ref{Qgeneral}) can be used to predict the equilibrium transport rate for arbitrary combinations of the density ratio $s$, Galileo number $Ga$ and Shields number $\Theta$ with $Ga\sqrt{s}\gtrsim81$ for non-rarefied drag. When compared with the simulations, these equations perform significantly better than the unified model of the cessation threshold and equilibrium transport rate of \citet{Pahtzetal21} (figure~\ref{PredictedMeasured}).
While the latter captures the $s^{1/3}$-dependence of $Q^+$, it fails to capture the $d^+$-dependence of $Q^+/s^{1/3}$ observed in the simulations.

\subsection{Effect of drag law and generalisation to drag in rarefied atmospheres} \label{GeneralizationDragRarefaction}
\begin{figure}
 \centerline{\includegraphics{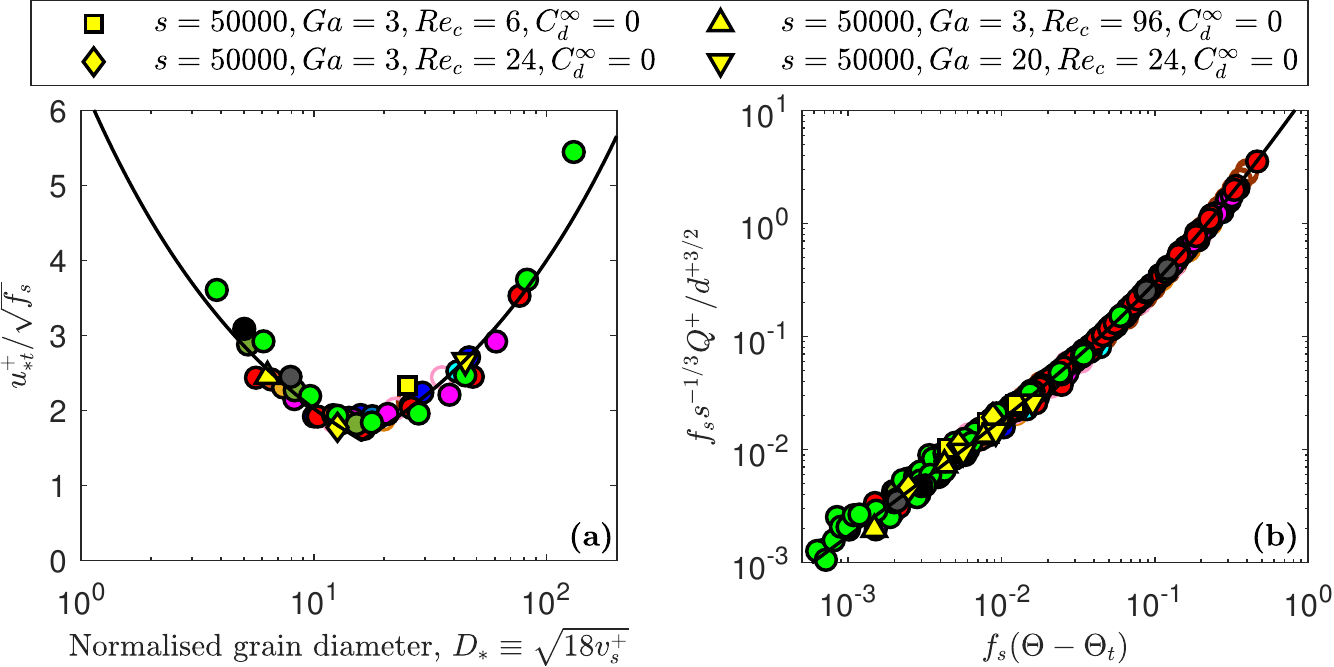}}
\caption{(a) Bedload-corrected rescaled cessation threshold shear velocity $u^+_{\ast t}/\sqrt{f_s}$ versus normalised median grain diameter, here redefined as $D_\ast\equiv\sqrt{18v_s^+}=\sqrt{24s/\Rey_c}d^+$. (b) Bedload-corrected normalised sediment transport rate $f_ss^{-1/3}Q^+/d^{+3/2}$ versus bedload-corrected Shields number in excess of the cessation threshold $f_s(\Theta-\Theta_t)$. Non-yellow symbols correspond to numerical simulations, carried out using the standard drag law parameters $\Rey_c=24$ and $C_d^\infty=0.5$, for various combinations of the density ratio $s$ and Galileo number $Ga$ (see table~\ref{SimulationPhaseSpace} and figure~\ref{PhaseSpace}) with $Ga\sqrt{s}>81$, and Shields number $\Theta$, where open and filled symbols indicate bedload and saltation conditions, respectively (see figure~\ref{PhaseSpace} for the definition). Yellow symbols correspond to additional simulations with modified drag law parameters as indicated in the legend.}
 \label{SettlingVelocityInfluence}
\end{figure}

The analysis in section~\ref{Proof} suggests that the normalised median grain diameter $D_\ast\equiv\sqrt{s}d^+$ in (\ref{Ut}) and (\ref{Utgeneral}) should be redefined as $D_\ast\equiv\sqrt{18v_s^+}=\sqrt{24s/\Rey_c}d^+$ (from (\ref{Drag})), which is equal to $\sqrt{s}d^+$ only in the case of non-rarefied drag ($\Rey_c=24$). To test this prediction as well as the effect of the form drag coefficient $C_d^\infty$, we carried out additional simulations using $C_d^\infty=0$ and $\Rey_c=[6,24,96]$ for a few saltation conditions. We find that these simulations, indeed, still satisfy (\ref{Utgeneral}) and therefore (\ref{Ut}) when the redefined $D_\ast$ is used (figure~\ref{SettlingVelocityInfluence}(a)).
They also still satisfy (\ref{Qgeneral}) and therefore (\ref{Q}), which are not affected by the redefinition of $D_\ast$ (figure~\ref{SettlingVelocityInfluence}(b)).

In rarefied atmospheres, the mean free path $\lambda$ of the air molecules becomes comparable to the median grain diameter $d$, or the Knudsen number $Kn\equiv\lambda/d=\sqrt{\pi k/2}s/(c^+d^+)$ \citep{Croweetal12}, with $c^+=c/(\tilde g\nu)^{1/3}$ the rescaled speed of sound and $k$ the adiabatic exponent, comparable to unity. This leads to a $Kn$-dependent correction $f_{Kn}\equiv1+Kn[2.49+0.84\exp(-1.74/Kn)]$ \citep{Croweetal12} of Stokes drag via $\Rey_c=24/f_{Kn}$. Note that typically $f_{Kn}\simeq1$ for $s\lesssim10^6$. Hence, the results in figure~\ref{SettlingVelocityInfluence} support that the following generalised definition of $D_\ast$ should be used for highly rarefied atmospheres ($s\gtrsim10^6$):
\begin{equation}
 D_\ast\equiv\sqrt{f_{Kn}}\sqrt{s}d^+. \label{Dgeneral}
\end{equation}

\section{Discussion} \label{DynamicThresholdMeasurements}
\subsection{Choice of dynamic-threshold measurements for evaluation of cessation threshold models}
Equilibrium saltation becomes intermittent below the continuous-transport threshold, characterised by alternating periods of equilibrium saltation and periods of rest \citep{MartinKok18}. The cessation threshold is therefore the wind strength at which equilibrium saltation would cease if extrapolated from the continuous-transport regime, that is, as the zero-point of equilibrium transport equations such as (\ref{taug}) or (\ref{Q}). It is also the threshold of intermittent saltation \citep{MartinKok18}. To evaluate the cessation threshold scaling law in (\ref{Ut}), we have therefore chosen exclusively measurements that either extrapolate continuous saltation in some manner to vanishing transport \citep{Cliftonetal06,Zhuetal19} or directly measure the cessation of intermittent saltation \citep{Bagnold37,Chepil45,Sugiuraetal98,MartinKok18,Zhuetal19}. Both methods require that equilibrium transport conditions can at least temporarily be established during the experiments \citep{Pahtzetal20a}, usually by feeding sufficient sediment when the test section is too short for transport to reach equilibrium. This requirement was probably not satisfied in all of the above-cited measurements. The snow drift wind tunnel by \citet{Cliftonetal06}, who did not feed snow at the tunnel entrance, was probably too short to establish equilibrium conditions for their beds of old and therefore cohesive snow, since cohesion can dramatically increase the fetch required to reach equilibrium \citep{Comolaetal19a}. For this reason, we have only compared with their data for freshly fallen snow.

Unfortunately, many other studies have not employed the same criteria when choosing measurements to evaluate their cessation threshold models \citep{ClaudinAndreotti06,Kok10b,Berzietal17,Andreottietal21,GunnJerolmack22}. This has largely been driven by the belief that there is only one dynamic threshold, implying that any kind of dynamic-threshold measurement is at least a proxy for the cessation threshold. However, we have presented evidence for the hypothesis that the continuous-transport threshold is a second kind of dynamic threshold with an underlying physics different from that of the cessation threshold \citep{PahtzDuran18a,Pahtzetal20a,Pahtzetal21}. An important example for a potential misinterpretation of measured dynamic thresholds as cessation thresholds is the study by \citet{Andreottietal21} for the following reasons.
\begin{enumerate}
 \item \citet{Andreottietal21}, who carried out their measurements in a pressurised-wind tunnel, explicitly mentioned that they were only able to establish equilibrium transport for air pressures relatively close to ambient pressure ($P\gtrsim30000~\mathrm{Pa}$) but not for the vast majority of studied pressure conditions (down to $P\approx200~\mathrm{Pa}$): `below $300~\mathrm{hPa}$ [the erosional zone] encompasses the entire bed.'
 \item \citet{Andreottietal21} explicitly defined threshold conditions `as the transition between saltation of groups of particles (bursts) to intermittent saltation of single particles (at high pressure) or no transport (at low pressure).' For high-pressure conditions, the so measured threshold is, by definition, larger than the cessation threshold \citep[i.e., the threshold of intermittent saltation;][]{MartinKok18}. For low-pressure conditions, the measurements are difficult to interpret due to the lack of equilibrium transport.
 \item \citet{Andreottietal21} accompanied their direct threshold measurements with indirect measurements obtained from extrapolating to vanishing transport. However, since they have not established equilibrium (for most pressure conditions), this extrapolation does not necessarily yield the cessation threshold.
 \item \citet{Pahtzetal21} hypothesised that the continuous-transport threshold is the smallest wind shear stress at which an average grain ejected by an impacting grain can be accelerated into a steady trajectory. A modification of their trajectory-based model based on this hypothesis captured the measurements by \citet{Andreottietal21}, suggesting that their employed experimental method yields a threshold akin to the continuous-transport threshold.
\end{enumerate}

The potential misinterpretation of the measurements by \citet{Andreottietal21} as cessation threshold measurements is highly relevant, since it led \citet{GunnJerolmack22} to introduce a Galileo number ($Ga$)-dependent viscous damping of grain--bed rebounds in their cessation threshold model in an attempt to capture these data. However, this rebound damping is the very reason for the very strong disagreement between their model and the here presented numerical data of the cessation threshold (section~\ref{CessationThresholdModelTest}). Note that, from a physical perspective, rebound damping should not depend on $Ga$ but on the Stokes number associated with the grain's impact velocity $St=\rho_p|\boldsymbol{v_\downarrow}|d/\mu$ \citep{Berzietal16,Berzietal17,Andreottietal21}, which is experimentally known to control the viscous damping of frontal grain collisions with a flat plate \citep{Gondretetal02}. Since typical values of $|\boldsymbol{v_\downarrow}|/\sqrt{\tilde gd}$ for Martian saltation are at the very least comparable to, if not much larger than, those for terrestrial saltation (because of $V_t\propto s^{1/3}\sqrt{\tilde gd}$, see section~\ref{PhysicsTransportRate}), and since $\rho_p$, $d$ and $\mu$ are of the same order of magnitude on Earth and Mars, typical values of $St$ on Mars are many orders of magnitude too large for viscous damping to play a meaningful role. In addition, even if there was a strong damping of frontal grain--plate collisions, this would not necessarily translate into a strong damping of grain--bed collisions. In fact, we previously reported only slight differences between DEM-RANS simulations of saltation for undamped (normal restitution coefficient $e=0.9$) and nearly fully damped ($e=0.01$) frontal grain--grain collisions \citep{PahtzDuran18a}. Even for $e=0.01$, grains can saltate in large hops on the order of $100d$ high \citep[Movie~S3 of][]{PahtzDuran18a}.

\subsection{Recommendations for how to reliably measure the saltation cessation threshold for low-pressure atmospheric conditions}
As described in the previous section, a reliable wind tunnel measurement of the cessation threshold for a given low-pressure atmospheric condition requires that equilibrium transport conditions can be established, at least temporarily. Since we are currently unable to predict with confidence the fetch distance saltation needs to reach equilibrium as a function of the atmospheric pressure, and since the required fetch could potentially be very large, it makes sense to design an experimental set-up that allows for adjustable sand feeding. However, this may be challenging given the closed-conduit nature of pressurised-wind tunnels. Once equilibrium transport is established in one way or another, we recommend to carry out measurements in the continuous-transport regime of the equilibrium transport rate $Q$ (or a proxy thereof) as a function of the shear velocity $u_\ast$ and then extrapolate these measurements to vanishing transport using $Q=c_1(u_\ast^2-u_{\ast t}^2)+c_2(u_\ast^2-u_{\ast t}^2)^2$ (consistent with (\ref{Q})), where $c_1$, $c_2$ and $u_{\ast t}$ are treated as fit parameters. The resulting cessation threshold $u_{\ast t}$ should be substantially smaller than the dynamic-transport threshold. In fact, for the terrestrial wind tunnel measurements by \citet{Creysselsetal09}, this extrapolation method yields the value $u_{\ast t}\simeq0.13~\mathrm{m/s}$ \citep{PahtzDuran20}, which is nearly a factor of $2$ smaller than the smallest wind shear velocity ($u_\ast\simeq0.24~\mathrm{m/s}$) for which \citet{Creysselsetal09} reported continuous equilibrium transport.

\section{Conclusions}
Guided by simulations with a well-established DEM-based numerical model \citep{Duranetal12} and existing experimental data, we have semi-empirically derived the scaling behaviours of the cessation threshold shear velocity $u_{\ast t}$ and rate $Q$ of equilibrium sediment transport across almost seven orders of the particle--fluid density ratio $s$, ranging from subaqueous transport ($s\approx2.65$) to aeolian transport in the highly rarefied atmosphere on Pluto ($s\approx10^7$). For saltation transport, occurring in planetary aeolian environments, they are
\begin{align}
 u_{\ast t}&=2.3(\tilde g\nu)^{1/3}\max\left[(D_\ast/18)^{-1/2},(D_\ast/18)^{1/2}\right], \label{utDimensional} \\
 Q&=1.7s^{1/3}\rho_p(d/\tilde g)^{1/2}(u_\ast^2-u_{\ast t}^2)+12s^{-2/3}\rho_p(\tilde g^3d)^{-1/2}(u_\ast^2-u_{\ast t}^2)^2, \label{QDimensional}
\end{align}
where $\rho_p$ is the particle density, $\nu$ the kinematic fluid viscosity, $\tilde g\equiv(1-1/s)g$ the buoyancy-reduced gravity, $d$ the median grain diameter and $D_\ast\equiv\sqrt{s}\tilde gd/(\tilde g\nu)^{2/3}$ its normalised value. In highly rarefied atmospheres ($s\gtrsim10^6$), $D_\ast$ is calculated by the more general (\ref{Dgeneral}), accounting for drag rarefaction effects. Put together, (\ref{utDimensional}) and (\ref{QDimensional}) constitute a simple means to make predictions of aeolian processes across a large range of planetary conditions.

The derivation of (\ref{utDimensional}) consists of a first-principle-based proof of the statement that $u_{\ast t}/(\tilde g\nu)^{1/3}$ is a function of only $D_\ast$ (section~\ref{Proof}). In contrast to existing cessation threshold models, this proof does not resort to coarse-graining the particle phase of the aeolian transport layer above the bed surface, but requires comparably much milder assumptions. Its arguably most critical underlying assumption is that scale-free boundary conditions describe the outcome of grain--bed collisions. The validation of the above statement with our extensive simulation data set therefore indicates that the characteristic velocity scale $\sqrt{\tilde gd}$ of grains ejected by the splash of an impacting grain plays no important role for the physics behind the cessation threshold. Instead, grain--bed rebounds and/or splash ejection events associated with the upper-tail end of the ejection velocity distribution are seemingly the physical processes that need to be considered.

The left and right term of the right-hand side of (\ref{QDimensional}) are consistent with the saltation limit and collisional limit, respectively, of the $Q$-scaling derived by \citet{PahtzDuran20}, with a threshold mean grain velocity scaling as $V_t\propto s^{1/3}\sqrt{\tilde gd}$. This scaling strongly deviates from the previous physical transport laws by \citet[][$V_t\propto\sqrt{\tilde gd}$]{UngarHaff87} and \citet[][$V_t\propto u_{\ast t}$]{Duranetal11}. For example, the law by \citet{UngarHaff87}, which has been adjusted to Earth conditions, underestimates the sediment transport rate for the simulated Martian conditions by a factor of about $5$. Only the recent model of \citet{Pahtzetal21} comes somewhat close to reproducing this scaling. It captures the $s^{1/3}$-dependence of $V_t$, but fails to capture its proportionality to $\sqrt{\tilde gd}$. This hints at a quite fundamental lack of understanding of the transport rate physics and calls for future studies on this issue.

For Martian atmospheric conditions, the cessation threshold values predicted by the numerical simulations and (\ref{utDimensional}) are much smaller than the recent dynamic-threshold measurements by \citet{Andreottietal21}. This is particularly odd given that both the numerical simulations and (\ref{utDimensional}) are in agreement with terrestrial experimental data. If the simulations were fundamentally wrong, one would expect them to fail for all conditions, not just for Martian conditions. In section~\ref{DynamicThresholdMeasurements}, we have therefore presented arguments for why the experimental methods used by \citet{Andreottietal21} may have yielded a threshold different from $u_{\ast t}$. This issue needs to be resolved in future studies, since knowing the `true' value of $u_{\ast t}$ is crucial for understanding the time evolution of Martian landscapes.

\backsection[Funding]{T.P. acknowledges support from the National Natural Science Foundation of China (no.~12272344). O.D. acknowledges support from the Texas A\&M Engineering Experiment Station.}

\backsection[Declaration of interests]{The authors report no conflict of interest.}

%\backsection[Data availability statement]{The data that support the findings of this study are openly available in}

%\backsection[Author ORCID]{Authors may include the ORCID identifers as follows.  F. Smith, https://orcid.org/0000-0001-2345-6789; B. Jones, https://orcid.org/0000-0009-8765-4321}

%\backsection[Author contributions]{}

%\appendix

%\section{}\label{appA}
 %This appendix contains sample equations in the JFM style. Please refer to the {\LaTeX} source file for examples of how to display such equations in your manuscript.

%\bibliographystyle{jfm}
%\bibliography{jfm}
%Use of the above commands will create a bibliography using the .bib file. Shown below is a bibliography built from individual items.
%\listofchanges
\bibliographystyle{jfm}
%\bibliography{model}

\end{document}